\documentstyle[12pt]{article}
\newcommand{\be}{\begin{equation}}
\newcommand{\ee}{\end{equation}}
\newcommand{\bea}{\begin{eqnarray}}
\newcommand{\eea}{\end{eqnarray}}

\newcommand{\pr}{\partial}
\newcommand{\g}{\gamma}
\newcommand{\lb}{\label}

\newcommand{\F}{\frac}

\newcommand{\X}{\cosh}

\newcommand{\5}{\left}
\newcommand{\6}{\right}

\newcommand{\al}{\alpha}

\newcommand{\rf}[1]{(\ref{#1})}
\newcommand{\non}{\nonumber}
\newcommand{\bold}{\load{\normalsize}{\bf}}
\begin{document}

\addtolength{\baselineskip}{0.20\baselineskip}

\hfill NORDITA-94/75 A

\hfill gr-qc/yymmnn

\hfill December 1994
\begin{center}

\vspace{24pt}

{{\Large \bf Quantum versions of Carlini-Miji\'{c} \\ wormholes }}

\end{center}

\vspace{12pt}

\begin{center}
{\sl A. Carlini$~^{\diamondsuit}$}
\footnote{Email: carlini@nbivax.nbi.dk}
{\sl D.H. Coule$~^{\ast}$}
\footnote{Email: coule@uctvax.uct.ac.za}
{\sl and D.M. Solomons$~^{\ast}$}
\footnote{Email: deon@maths.uct.ac.za}

\vspace{12pt}

$~^{\diamondsuit}$NORDITA, Blegdamsvej 17, DK-2100 Copenhagen \O, Denmark

$~^{\ast}$Department of Applied Mathematics, University of Cape Town,

Rondebosch 7700, South Africa

\vspace{12pt}
{\bf Abstract}
\end{center}

\bigskip
\bigskip
\bigskip

We consider the quantum analogues of wormholes obtained by Carlini and Miji\'c
(CM), who analytically continued closed universe models. To obtain
wormholes when the strong energy condition ($\gamma>2/3$) is satisfied,
we are able to simplify the Wheeler-DeWitt (WDW) equation by using an
equivalent scalar potential which is a function of the scale factor. Such
wormholes are found to be consistent with the Hawking-Page (HP) conjecture
for quantum wormholes as solutions of the WDW equation.

In addition to the CM type wormholes, for a scalar field realization of
the potential in the WDW equation we also obtain quantum
wormholes when the strong energy condition is violated. This violation can
be up to an arbitrary large distance from the wormhole throat, before
the violation eventually has to be relaxed in order to have a flat
Euclidean space time. These results give support to the claim of HP that
wormhole solutions are a fairly general property of the WDW equation.
However, by allowing such solutions one might be precluding other more
important properties such as a Lorentzian behaviour and a possible inflationary
earlier stage of our universe.

\vfill

\newpage

\section{Introduction}

There is at present much interest in the use of wormholes. This is
because they could affect the value of coupling constants in our
universe. For the
 cosmological constant $\Lambda$ the effect of wormholes is to
make $\Lambda$ a dynamical variable given by a distribution function
$P(\Lambda)$. If such a distribution is peaked around
$\Lambda\rightarrow 0$, it could help explain why the cosmological constant
is zero in our universe \cite{col}.

Because wormholes contain information that is not necessarily accessible
to outside observers, the quantum state appears to be mixed. This is
reminiscent of black holes, and so wormholes might play a role in
understanding Hawking radiation from black holes and their resulting
evaporation \cite{haw}.

Originally, classical wormholes were constructed using axions. Without
reviewing all the cases that have since been found, we consider some
of their general features. It is conjectured that wormholes are
possible if the Ricci tensor can have a negative eigenvalue \cite{shoen}, of
course for
a given convention. For a minimally coupled real scalar field
the Ricci tensor is given by
\begin{equation}
R_{\mu\nu}=\partial_{\mu}\phi\partial_{\nu}\phi + g_{\mu\nu}V(\phi)
\lb{a}
\end{equation}
Because in Euclidean space $g_{\mu\nu}=(++++), R_{\mu\nu}>0 $ for
$V(\phi)>0$ and wormholes do not occur for, say, a massive scalar
field with potential $V(\phi)=m^2\phi^2/2$. However, if we consider an
imaginary
scalar field, i.e. let $\phi\rightarrow i\phi$, then $R_{\mu\nu}$ can
become negative. For an imaginary massive scalar field we now get
\begin{equation}
R_{\mu\nu}=-\partial_{\mu}\phi\partial_{\nu}\phi -g_{\mu\nu}m^2\phi^2
\lb{b}
\end{equation}
and a wormhole solution should now be possible. For this imaginary massive
scalar field the wormhole solution is given in  ref. \cite{imag}. However a
similar reasoning is valid with other classical wormholes.

Recently, a
paper \cite{cots} has appeared which seemingly contradicts this reasoning: it
is
claimed that a real scalar field with a non-negative potential can give
a wormhole. The authors then simulate this potential using higher order
corrections to the Lagrangian for gravity. Instead they should have
found that a negative potential for the scalar field
was necessary, but that this can
be simulated by higher order corrections with ``wrong signs'' chosen. This
would then have agreed with the reasoning for why the $R+\epsilon
R^2$ wormhole is possible when the wrong sign of $\epsilon$ is chosen.
Wrong in the sense that the theory would be unstable in Lorentzian
space-this was previously explained in ref. \cite{cou1}.

Instead of looking for special matter sources that give wormholes,
Carlini and Miji\'{c} (CM) \cite{cm} considered an analytic
continuation of closed
Friedman-Robertson-Walker (FRW) universes.  For a perfect fluid equation
of state  $p=(\gamma-1)\rho$, ($p$ and $\rho$ are the pressure and
energy density respectively)
closed universes require that the strong energy condition
be satisfied, i.e. $\gamma>2/3$.  By using an adaptation of an approach
of Ellis and Madsen (EM) \cite{em},
they worked with a scalar field model with a
potential $V(\phi)$ which
simulates a certain value of $\gamma$. Or one might say that, by
altering the parameter $\gamma$, the slope of the scalar field potential can
be altered. If the slope is too shallow the strong energy condition
is violated and we would not expect wormholes to be possible. By being
able to alter the slope of the scalar potential it gives a
 more systematic way
 of investigating when wormhole solutions can occur
than by just choosing certain
potentials, for example a massive scalar field $V(\phi)=m^2\phi^2$.

Originally, CM used an asymmetric continuation to find the
corresponding wormhole from a closed universe. The matter and
gravitational parts were rotated in different directions as the
transition to the Euclidean regime was done. This is similar to
the procedure of Linde \cite{lin} for obtaining a positive gravitational
action $S$ which
could describe the quantum creation of the universe through
tunneling, with probability $\sim \exp(-S)$. Later this continuation
was extended in ref. \cite{c1} to also include the scalar field $\phi$.
These methods have the
advantage of avoiding discontinuities in the scalar potential and
the kinetic energy at the junction point. However, the nature of
these continuations ensures that the Ricci tensor has negative
eigenvalues in the Euclidean regime. This means that these wormholes
are still rather restrictive, but this aspect has been ``hidden'' in the form
of the continuation. We note in passing that some wormholes
(e.g. $R+\epsilon R^2$) are
more closely related to the analytic continuation of bounces \cite{cou1}:
when  the singularity  in open
Lorentzian  $k=-1$ universes is avoided.

In a different vein, because the number of known
classical wormhole solutions had appeared so restrictive,
Hawking and Page (HP) considered that solutions of the
Wheeler-DeWitt (WDW) equation could more generally represent
wormholes \cite{hp}. They hoped that all reasonable matter sources would
have the possibility of realizing quantum wormholes.
For such wormholes they suggested that the wavefunction $\Psi$ should
decay exponentially for large scale factors $a$ so as to
represent Euclidean space, and that $\Psi$ be well behaved as $a\rightarrow
0$, so that no singularities are present.

 Because the WDW equation is independent
of the lapse, the Euclidean regime is already included
in the formalism. We hope to find that the quantum versions of
the CM wormholes are automatically included in
the solutions of the WDW equation, so avoiding the ad-hoc
continuation required for the classical wormholes. There is
some hope of expecting that this is possible: the WDW equation
has both Hartle-Hawking (HH) \cite{hh} and Tunneling \cite{tun}
boundary condition
solutions. As already mentioned, when dealing with the classical
action the Tunneling case corresponds to an asymmetric continuation.
This is
rather analogous to what will be required to give quantum CM wormhole
solutions.

As well as finding the quantum analogues of wormholes that occur
when the strong energy condition is satisfied, we  also wish
to explore if wormholes are even more general. We will do this by
using the potential that simulates a certain value of $\gamma$ .
Hawking and Page constructed quantum wormholes for a real massive scalar
field and a $\phi^n$ potential. But, depending on where the scalar field
is on such potentials, the effective value of $\gamma$ can be anywhere in
the range $0\leq\gamma\leq2$ and it was not clear if the
strong energy condition was violated in such cases.
 If wormholes can also  occur when the
strong energy condition is violated, this would greatly extend the
viability of wormholes in mediating processes where this
condition is  probably violated, for example in black hole evaporation.

\section{The classical model}
We first review the classical closed universe models from which we
are going to derive their corresponding WDW equations.
For a more detailed description of the features of the classical CM
wormholes, we refer to the original results presented in ref. \cite{cm} and
to the Appendix.
First we take a bulk matter source with a perfect fluid equation
of state $p=(\gamma-1)\rho$ and work in a Lorentzian
FRW ansatz
\be
ds^2=\hat\sigma^2[-N^2dt^2+a^2d\Omega_3^2]=\hat\sigma^2[-a^{4-3\g}dt^2+a^2d
\Omega_3^2]
\lb{x}
\ee
where the lapse is chosen to aid calculations as in ref. \cite{cm}, and
$\hat\sigma^2=2G/ 3\pi$.

The scale factor $a$ is given by
\be
a=a_0\left [1-\left({3\g-2\over 2}\right )^2{t^2\over a_0^{3\g-2}}
\right ]^{1/(3\g-2)}
\lb{ff11}
\ee
with $a_0$ an arbitrary constant which is the maximum size of the
FRW universe when $\gamma>2/3$.
 Using the same approach as in EM, one can convert
to a scalar field $\phi$ whose trajectory is given by
\be
\phi-\phi_0=\frac{\sqrt{2\gamma}}{3\gamma-2}\tanh^{-1}\left [
\frac{3\gamma-2}{2a_0^{(3\gamma-2)/2}}t\right ]
\lb{f111}
\ee
 The  classical potential for the scalar field ($\phi_0=0$) is
\be
V(\phi)=\Omega\cosh^{2n}\lambda\phi
\lb{ff15}
\ee
where we have defined
\be
\lambda=\frac{3\gamma-2}{\sqrt{2\gamma}}~~~,~~~n={3\g\over 3\g-2}~~~,
{}~~~\Omega={(2-\g)\over 2a_0^2}
\label{param}
\ee
Note the difference in the previous expressions with their Euclidean
counterparts in ref. \cite{cm}.
These expressions are valid in a closed universe,
but analogous expressions could be obtained for $k=-1$.

The classical CM wormhole is obtained by analytically continuing
the time variable as $t\rightarrow -it$ in the
closed universe eq. \rf{ff11}.
These classical wormholes appear to have a nontrivial potential term,
and so do not possess any conserved charge. The fact that they
exhibit a periodicity in the Euclidean time
can be interpreted as evidence that wormholes of size $a_0$ have a
finite temperature $T\sim 1/a_0$ \cite{cm}.
Inclusion of a small bare cosmological constant was also considered in
ref. \cite{c1}.
Furthermore, the analytic continuation to the Euclidean regime was shown
to be consistent with the reality of the Euclidean path integral at
one-loop \cite{c2}.

By using eqs. \rf{ff11} and \rf{f111},
the scalar potential can be written as a
function
of the scale factor
\be
V(\phi)\equiv V(a)=\frac{V_m}{a^{3\gamma}}
\lb{fshell}
\ee
where the constant $V_m=\Omega a_0^{3\gamma}$.

Before we go on to consider the quantum versions of these wormholes,
let us make two important comments about what has been done.

 a) the
equivalence between the potentials $V(a)$ and $V(\phi)$ is only valid
``on shell'' when all the classical equations are satisfied. The
equivalence needs not to occur in the quantum theory when only the WDW equation
has to be satisfied (see ref. \cite{teit}).
In such cases the potential $V(\phi)$ will contain
 possibilities of extra solutions since it is not constrained by an
additional classical  equation.

b) Although the potentials $V(\phi)$ and $V(a)$ both simulate a certain
value of $\gamma$, they only do this for a finite time as the scalar
field `rolls down' the potential. Eventually the simulation will break
down and either the kinetic or potential energy will dominate, i.e. $\gamma$
will tend to 2 or 0.
 In this case the potentials do not have the same ground state
since, when $\phi=0$, the $V(\phi)$ potential behaves like a cosmological
constant. In the classical theory this difference in  the ground states
could be set to only occur at an asymptotically large time in the future.
But it is well known that the WDW equation has no explicit time dependence
and so this aspect of the ground state cannot be isolated in this way.
It can be treated by introducing a second scalar field
(or equivalently using a complex scalar field) as was also
done in ref. \cite{ell1} when extending the EM approach. This problem
does not occur, in any case, when $\gamma<2/3$, as the two potentials neither
have a cosmological constant  ground state.
Strictly speaking, all the simulations break down in the limit $a\rightarrow
\infty$, since the effective $\gamma$ is related to the ``roll down''
along the potential.
On reaching the minimum, the potential energy is lost and $\gamma$ will tend
to $2$, the kinetic dominated case.
Because of this limitation we can not claim that
 wormholes (with a certain value of $\gamma$) occur in the limit $a\rightarrow
\infty$, but only
 up to a finite scale factor $a_{max}$. However, this size  can be made
much greater than the Planck scale, and for all practicality  this is not
a serious limitation.

\section{The ``on shell" $V(a)$ matter model}
We consider solutions of the WDW equation, see e.g. ref. \cite{hall}
\be
\left ( \frac{\partial}{\partial a^2}+\frac{p}{a}\frac{\partial}
{\partial a}-\frac{1}{a^2}\frac{\partial }{\partial \phi^2} -ka^2
+a^4V(\phi) \right ) \Psi(a,\phi)=0
\lb{dewitt}
\ee
where $p$ is a factor ordering correction and $k$ the spatial curvature,
$k=\pm 1,0$.

We will start our quantum analysis of the CM wormholes
for the case in which the matter content of the theory
is given by the ``on shell'' potential $V(a)$. As mentioned, this will
 not exhaust all the possible solutions since we
could use the more general  potential $V(\phi)$. However we
first try to find the wanted solutions in this way since they are a
subset of all the solutions  present when  using $V(\phi)$.
For the potential $V(a)$ the WDW equation is
separable, and simplifies to:
\be
\left ( a^2\frac{d^2}{da^2}+pa\frac{d}{da}+q^2+V_ma^{6-3\gamma}-ka^4
\right ) \Psi(a)=0
\lb{treuno}
\ee
\be
\left ( \frac{d^2}{d\phi^2}+q^2\right ) \Psi(\phi)=0
\lb{tredue}
\ee
with $q$ the separation constant.

If we had used a relativistic perfect fluid, the corresponding WDW equation
would have taken the form (see, e.g., ref. \cite{ryan})
\be
\left ( a^2{d^2\over da^2}+pa{d\over da}+\rho_0a^{6-3\g}
-ka^4\right ) \Psi(a)=0
\lb{f71}
\ee
where we have substituted from the conservation equation the energy
density $\rho=\rho_0/a^{3\gamma}$, with $\rho_0=a_0^{3\g-2}$.
If possible we will
 use this simpler equation because all
its solutions are contained as  special cases of eqs. \rf{treuno} and
\rf{tredue},
i.e. setting $q=0$ and  taking the redundant $\phi$ solution
 $\Psi(\phi)=$constant .

 We must impose some boundary condition in order to select
(if any) the solutions of the WDW which represent asymptotically
Euclidean (AE) wormholes.
Here we follow the proposal of HP:

\* {\bf (a)}
$\Psi$ should decay exponentially as the radius $a \rightarrow \infty$,
like $\exp(-\frac{1}{2}a^2)$;

\* {\bf (b)} $\Psi$ should be well-behaved (regular) at the origin.

This is because $\Psi$ should represent Euclidean space for large $a$
and there should be no singularities as $a\rightarrow 0$. It should also
not have any divergences due to the matter content, which would correspond
to singularities.

We should point out that throughout this paper we are only interested
in the possible existence of such wormhole solutions and set
arbitrary coefficients accordingly. This choice depends on the
boundary conditions applied. We will later discuss their relation
to the more usually applied boundary conditions of HH
and the Tunneling ones.

We can get a preliminary idea as to when an Euclidean domain
occurs
at large $a$ by studying the WDW equation \rf{f71} as an ordinary
Schr\"odinger equation (for the factor ordering $p=0$)
\be
\left({d^2\over da^2}+U(a)\right )\Psi(a)=0
\lb{f71a}
\ee
which represents the motion of a `particle' of unit mass and zero energy
in the potential $U(a)=\rho_oa^{4-3\g}-ka^2$.
When $U>0$ the wave function for large $a$ is oscillating, implying
the existence of a Lorentzian phase (see, e.g., ref. \cite{hall}).
Therefore, in order to have an AE wormhole, it is necessary that, at
least, $U<0$.
Returning to eq. \rf{treuno}, and setting the unimportant in
this regard factor $p=0$, this occurs for
\be
V_ma^{4-3\gamma}-ka^2<0
\lb{f72}
\ee
Therefore for the usual case of a positive potential ($V_m>0$, $\g<2$) we
require $2>4-3\gamma$, i.e. $\gamma>2/3$ and $k=1$ for such behaviour.
If we had a negative potential $(V_m<0$) then this would be reversed:
$\gamma<2/3$ and $k=\pm1$ can give such a solution. This is an example
that shows how $k=1$ is not strictly necessary to obtain wormhole
solutions - at least in the CM and HP sense.
We have ignored the term involving the separation constant $q^2/a^2$
which would only change this argument as $a$ becomes small.

This strong energy condition is the same as that obtained by CM for
the occurrence of wormhole solutions.
This condition is stronger than (but consistent with) the result of
ref. \cite{kp} that quantum wormholes are incompatible
with a cosmological constant.

A quantum wormhole also requires a suitable behaviour for small $a$. As $
a\rightarrow 0$ we can ignore the $ka^4$ term, since $4>6-3\gamma$ when
$\gamma>2/3$. In this
case the WDW equation \rf{treuno} simplifies to a Bessel equation
with solution
\be
\Psi(a)\simeq a^{1-p\over 2}\left [a_1J_{i\tilde q}\left ( {2a^{3-3\gamma/2}
\sqrt{V_m}\over 3(2-\g)}\right)+a_2
Y_{i\tilde q}\left ({2a^{3-3\gamma/2}\sqrt{V_m}\over 3(2-\g)}\right)\right ]
\lb{f75}
\ee
where $\tilde q=[4q^2-(1-p)^2]^{1/2}/3(2-\g)$.
We have included the separation constant in the solution in order to outline
a potential problem it introduces: that of an infinite oscillation as
$a\rightarrow 0$. Note that this problem is absent for a perfect fluid matter
source when $q=0$, but needs to be taken care of when using a scalar
field source. We next outline this problem and its resolution.

Using the asymptote  $J_{\nu}(z)\sim z^{\nu}$, as $z\rightarrow 0$ (see
ref. \cite{abr}), enables
the solution for the ordering factor $p=1$ to be expressed as\be
\Psi(a)\sim \exp\left [iq \ln a\right ]
\lb{f76}
\ee
The other Bessel function would simply have a $(-)$ sign in the exponent
of eq. \rf{f76} since $Y_{\nu}(z)\sim -z^{-\nu}$. The foregoing argument would
proceed in the same fashion.

The problem now is that as $a\rightarrow 0$ the $\ln a\rightarrow -\infty$
causes infinite oscillations to occur, and the wavefunction cannot be regarded
as a wormhole in its present form as the oscillations represent a
singularity \cite{hp}. The full solution is, however,
\be
\Psi(a,\phi)\sim \exp{iq[\ln a+\phi]}
\lb{f79}
\ee
By integrating over the separation constant we can eliminate
this singularity at the origin. This is the same as performing a Fourier
integral \cite{zhuk} or like constructing a wave-packet solution
\cite{kief}.
Now the integral
\be
\Psi\sim \int \exp\left \{iq\left [ \ln a +\phi
\right ] \right \}dq
\lb{f80}
\ee
is of the form $\int \exp(ixt)dt$ which, by means of the Riemann-
Lebesgue lemma (see, e.g., ref. \cite{bend}),
tends to zero as $x\rightarrow \infty$.
The wavefunction is now, in effect, damped as $a\rightarrow0$ or $\phi
\rightarrow \infty$.
This now satisfies the regularity boundary condition {\bf (b)} of HP.
 Note that there is still a possible divergence due to the
  factor ordering term $\sim a^{1/2-p/2}$ when $p>1$. It has
been suggested \cite{kim} that even when this divergence is present it
should  still be considered to satisfy the  HP condition for
wormholes. Otherwise we would conclude that for certain factor ordering
values $(p>1)$ wormholes are prevented due to this divergence.
\subsection{Two examples}

We conclude this first section by giving two explicit examples of
solutions of the WDW equation for the ``on shell'' potential $V(a)$ of a
perfect
fluid matter model. Although the WDW equation has been simplified
to an ordinary differential equation, it is still not straightforward
to obtain analytic solutions for all $\gamma$. One could proceed by
finding approximate WKB solutions. But instead we consider two cases
of $\gamma$: one satisfying and one violating the strong energy condition.
This enables us to emphasize properties that any solution in the range
$0\leq\gamma\leq 2$ will have.

\paragraph{~~$\bullet$~~\bold $\bf{\g =4/3}$ }
\indent

This example classically represents a radiation dominated FRW geometry or
that of a conformally coupled field.
In this case the WDW equation \rf{f71}, for the factor ordering $p=0$,
 can be rewritten as
\be
\left\{{d^2\over da^2}+ a^2_0-a^2\right\}\Psi(a)=0
\lb{f78}
\ee
which can be thought of as the Schr\"odinger equation for a harmonic oscillator
with energy $a^2_0$.
The general solution of eq. \rf{f78} can be expressed as a linear superposition
of harmonic wave functions as
\be
\Psi(a)=\sum_nc_n\exp(-a^2/2)H_n(a)
\lb{f78a}
\ee
where $H_n$ are the Hermite polynomials.
Such wave functions are regular at the origin and exponentially damped
at infinity, and according to the HP boundary conditions they represent
quantum AE wormholes, see fig. (1).
The minimum `throat' of the wormholes is quantized, $a_0=\sqrt{2n+1}$.

\paragraph{~~$\bullet$~~\bold $\bf{\g =0}$ }
\indent

This example is equivalent to having a cosmological constant as the
matter source.
In this case the WDW equation \rf{f71}, for the factor ordering $p=0$,
 can be rewritten as
\be
\left\{{d^2\over da^2}+{a^4\over a_0^2}-a^2\right\}\Psi(a)=0
\lb{f77}
\ee
This is reminiscent of the Schr\"odinger equation for an anharmonic
oscillator, and it can be exactly solved by a suitable redefinition
of variables in terms of Airy functions as
\be
\Psi(a)=b_1~ Ai\left [2^{-2/3}\left(1-{a^2\over a_0^2}\right)\right]+
b_2~ Bi\left [2^{-2/3}\left(1-{a^2\over a_0^2}\right)\right]
\lb{f77a}
\ee
The wave function is oscillating for large $a$ (see fig. (2)),
and clearly does not resemble an AE wormhole (as we expected, since
$\g<2/3$).
Actually, it may be thought as representing
the quantum nucleation at the radius
$a_0$ of an expanding Lorentzian inflationary universe.

\section{The scalar field model}
\subsection{The general $V(\phi)$ potential case }
So far we have obtained quantum wormholes with the restriction that the
strong energy condition is satisfied.
We next  extend our analysis of the quantum solutions by studying
the WDW equation with the scalar potential  $V(\phi)$.
This potential gives the full  possibility of solutions and so might
also give quantum wormholes when the strong energy condition is
violated, i.e. when $\gamma<2/3$.

 This potential is plotted for various values of the parameter $\gamma$ in
figs. (3-5). For $\gamma>2/3$ a positive minimum occurs when $\phi=0$. At this
point the simulation of $\gamma$ breaks down and the actual value of
$\gamma$ becomes zero - it behaves as a cosmological constant.
 In order to analyse this region we would have
to prevent the potential behaving like a cosmological constant. This can be
done by adding an extra field to enable the ground state to correspond to
 $V(\phi)=0$. We later outline how this can proceed.

 However, we already know that wormholes occur for $\gamma>2/3$
and so we can restrict our attention to when $\gamma<2/3$. In this case the
potential $V(\phi)$ has a zero ground state $V(\phi)\rightarrow 0$ as
$\vert\phi\vert\rightarrow \infty$ and so there is no problem of a
cosmological constant occurring. We have to ensure that
the simulation of  $\gamma<2/3$ does not break down  before the wormhole
has an arbitrary large size $a$. As previously mentioned, in the limit
$a\rightarrow \infty$, we require asymptotically flat Euclidean space,
i.e. $V(\phi)=0$.

\subsubsection{Separating the WDW}
We now consider the WDW equation for the case of the $V(\phi)$
potential when the strong energy condition  is violated.
As already mentioned, the only region when we can hope to find AE
wormhole solutions is the one where the potential $V(\phi)$ is
approaching its minimum, i.e. the region of superspace where
$\vert \phi \vert \rightarrow \infty$. Taking $\phi \simeq 0$ would
imply expanding the potential of eq. \rf{ff15} around an (unstable)
cosmological
constant value, which HP showed is inconsistent with the existence of
AE wormholes.

Let us therefore try to separate the WDW eq. (9) by  introducing
 a new variable $\eta$ as
\be
\eta = f(\phi) a^3
\label{e1}
\ee
for some function $f$ of the scalar field $\phi$.
Defining the quantity $F$ such that
\be
F^{-1} = \frac{1}{f}\frac{\pr f}{\pr\phi}
\label{e2}
\ee
the WDW equation \rf{dewitt} may be then expressed as
\[
F^2\left[a^2\frac{\pr^2}{\pr a^2} + pa\frac{\pr}{\pr a} -
a^4\right]\Psi(a, \eta) =\]
\be
\left[ \left(\eta^2\frac{\pr^2}{\pr \eta^2} + (1 - F')\eta \frac{\pr}{\pr
\eta}\right) - a^6VF^2\right]\Psi(a, \eta)
\label{e5}
\ee
By inspection it seems useful to assume that the variable coefficient
\be
\frac{1 - F'}{F^2}
\lb{e3}
\ee
and the term
\be
-a^6V(\phi)
\lb{e4}
\ee
are proportional, i.e. that
\be
\frac{1 - F'}{F^2} = \frac{1}{\delta}f^{-2}V(\phi)
\label{e24}
\ee
where $\delta$ is a constant of proportionality.
It is then easy to check, using eq. \rf{e2},
that eq.~\rf{e24} reduces to the second order differential equation
\be
f(\phi)f''(\phi) = \frac{1}{\delta}V(\phi)
\label{e26}
\ee
The next step is to observe that there exists a function
(say $y$) that matches the behaviour of $V(\phi)$, and satisfies the
differential equation
\be
yy'' = \alpha y'^2 + \beta yy' + \rho y^2
\label{e27}
\ee
with constant coefficients $\alpha, \beta$ and $\rho$.
It is easy to verify that a possible solution of eq. \rf{e27} is
given by
\be
y(z) = e^{hz}\cosh^{\frac{1}{1 - \alpha}}{k(\alpha - 1)z}
\label{e30}
\ee
where
\bea
h & = & -\frac{\beta}{2(\alpha - 1)}\label{e28}\\
k & = & \frac{1}{2(\alpha - 1)}\sqrt{\beta^2 - 4\rho(\alpha - 1)}\label{e29}
\eea
Moreover, direct substitution into eq.~\rf{e27} results in
\be
yy'' = y^2[ \alpha h^2 + \beta h + \rho - (\beta + 2\alpha h)k
\tanh{k(\alpha - 1)z} + \alpha k^2\tanh^2{k(\alpha - 1)z}]
\label{e31}
\ee
To simplify things, we set $\beta= 0$ (and therefore $h =
0$), in eqs.~\rf{e27}-\rf{e31}. Then eq.~\rf{e31} simplifies to
\bea
yy'' & = & y^2[\rho + \alpha k^2\tanh^2{k(\alpha - 1)z}]
\non \\
     & = & y^2[k^2 -\frac{\alpha k^2}{\cosh^2{k(\alpha - 1)z}}]
\label{ee34}
\eea

The idea now is to consider the ansatz
\be
\cosh^2{k(\alpha - 1)z} \gg \alpha
\label{ee35}
\ee
for which eq.~\rf{ee34} becomes
\bea
yy'' & \approx & k^2y^2
\non \\
    &    =   &  k^2\cosh^{2\over 1-\al}k(\al-1)z
\label{ee37}
\eea
Compare now eq. \rf{e26} and eq. \rf{ee37}:
for the potential given by eq. \rf{ff15} we can, in fact, identify
\be
y = f(\phi)
\label{e38}
\ee
provided that we also equate the arguments of the potential~\rf{ff15}
and of the cosh-function in eq.~\rf{e30}, i.e. if we take
\be
\lambda = k(\alpha - 1)~~~~~~,~~~~~~z=\phi
\label{e39}
\ee
and we impose the further conditions
\be
{2\over 1-\al}=2n~~~,~~~k^2={\Omega\over \delta}
\lb{e39b}
\ee
In other words, we can write
\be
ff'' \approx  k^2f^2
\label{e41}
\ee
It is a matter of a simple algebra to show that equations~\rf{e39} and
\rf{e39b}, together with parameters
$\lambda , n$ and $\Omega$ given by eq. \rf{param}, imply that
\bea
\alpha &=& \frac{2}{3\gamma}
\non \\
\rho&=& \frac{3(3\g-2)}{2}
\non \\
\delta  &=& \frac{2-\g}{9\g a^2_0}
\label{ee59}
\eea

We find comfort in the fact that the ansatz \rf{ee35} is
satisfied in the interesting region $\vert\phi\vert\rightarrow \infty$.

Moreover, noting that in this case we have
\be
F^{-2}=k^2\tanh^2\lambda\phi \simeq
k^2 \simeq \frac{1 - F'}{F^2}
\label{ee44}
\ee
we can finally separate eq. \rf{e5} as
\be
\left[a^2\frac{d^2}{d a^2} + pa\frac{d}{d a} -a^4 +\nu
a^{2\xi}\right]\psi(a) =
0
\label{ee46}
\ee
and
\be
\left[ \eta^2\frac{d^2}{d \eta^2}
+ \eta{{d}\over {d\eta}} - \delta\eta^2+{\nu\over
k^2}\right]
\Phi(\eta) = 0
\label{ee47}
\ee
for a separation constant $\nu$ and with $\xi=3(3\g-2)/ 2(3\g-1)$.

The general solutions of eqs. \rf{ee46} and \rf{ee47} can be easily
expressed in terms of Bessel functions (see, e.g., ref. \cite{abr}).
In particular, the wave function $\psi(a)$ which is bounded as
$a\rightarrow \infty$ is
\be
\psi(a)= a^{1-p\over 2}K_{{1\over 2}\sqrt{{(1-p)^2\over 4}-\nu}}
\left ({a^2\over 2}\right )
\lb{ff53}
\ee
while the wave function $\Phi(\eta)$ for $\g<2$ is
\be
\Phi(\eta)= f_1K_{\sqrt{-\nu}\over \vert k\vert}(\sqrt{\delta}\eta )+f_2
I_{\sqrt{-\nu}\over \vert k\vert}(\sqrt{\delta}\eta )
\lb{ff54}
\ee

We now explicitly study the solutions of the eqs.
\rf{ee46}-\rf{ee47} in the limit of large $\vert\phi\vert$
for the case $\g<2/3$.

\paragraph{~~$\bullet$~~\bold ${\bf 0<\g<2/3}$ }
\indent

\subparagraph{~~$\diamond$~~${a\rightarrow 0}$ }
\indent

Both in the limit of small and large scale factors, the $a$ dependent part
of the wave function is given by the modified Bessel function
$K$ of eq. \rf{ff53}.

Turning to the $\eta$ dependent part of the solution, we remember that
the coordinate $\eta$ of eq. \rf{e1} reads as
\be
\eta =a^3\cosh^{2n}\lambda\phi\sim c_0a^3e^{\lambda_0\vert\phi\vert}
\lb{gg5}
\ee
in the limit of $\vert\phi\vert\rightarrow \infty$, where we have
introduced $\lambda_0=3\sqrt{\g/2}~\mbox{sign}(3\g-2)$ and $c_0=\exp
[-6\g/(3\g-2)\ln 2]$.
In the case of small $a$, since $\lambda_0 < 0$ for $\g<2/3$, the only
possibility is that $\eta\rightarrow 0$.

Therefore, combining the solutions \rf{ff53} and \rf{ff54} in the
limit of small arguments, we obtain for the asymptotic form of the global
wave function (for $p=1$):
\be
\Psi(a, \phi) \simeq a^{-\sqrt{-\nu}}\left [ g_1a^{-{3\sqrt{-\nu}\over
\vert k\vert}}e^{-{\lambda_0\vert\phi\vert\sqrt{-\nu}\over \vert
k\vert}}+g_2a^{{3\sqrt{-\nu}\over
\vert k\vert}}e^{{\lambda_0\vert\phi\vert\sqrt{-\nu}\over \vert
k\vert}}\right ]
\lb{gg1}
\ee
This wave function is regular and satisfies the boundary condition {\bf
(b)} of HP for $g_1=0$.

\subparagraph{~~$\diamond$~~${a\rightarrow \infty}$ }
\indent

In the limit $a\rightarrow \infty$, the two factors in the coordinate
$\eta$ compete one against the other and we have two possibilities
depending on the scaling: either $\eta\rightarrow 0$ or $\eta\rightarrow
\infty$.

In the case $a^3e^{\lambda_0\vert\phi\vert}\rightarrow 0$, the global
wave function (combination of eqs. \rf{ff53}-\rf{ff54}) has the
following asymptotics for large $a$:
\be
\Psi(a, \phi) \simeq a^{-{(1+p)\over 2}}
\left [g_3a^{-{3\sqrt{-\nu}\over
\vert k\vert}}e^{-{\lambda_0\vert\phi\vert\sqrt{-\nu}\over \vert
k\vert}}+g_4a^{{3\sqrt{-\nu}\over
\vert k\vert}}e^{{\lambda_0\vert\phi\vert\sqrt{-\nu}\over \vert
k\vert}}\right ]e^{-a^2/2}
\lb{gg2}
\ee
This is clearly consistent with the boundary condition {\bf (a)} of HP for
the existence of wormholes.

Similarly, in the case $a^3e^{\lambda_0\vert\phi\vert}\rightarrow \infty$,
the global wave function has the asymptotics
\be
\Psi(a, \phi) \simeq a^{-{(4+p)\over 2}}
e^{-{\lambda_0\vert\phi\vert\over 2}}\biggl [g_5e^{-{a^2\over 2}\left
[1+c_1ae^{\lambda_0\vert\phi\vert}\right ]}
+g_6 e^{-{a^2\over 2}\left [1-c_1ae^{\lambda_0\vert\phi\vert }\right ]}
\biggr ]
\lb{gg3}
\ee
where $c_1=2\sqrt{\delta}c_0$.
This can be AE in the sense of HP if we assume the further scaling
$ae^{\lambda_0\vert\phi\vert}\rightarrow 0$.

In conclusion, we find that the Lorentzian potential $V(\phi)$ given
by eq. \rf{ff15} is consistent with the existence of quantum AE wormholes
according to HP in the case $\g<2/3$ and
$\vert\phi\vert\rightarrow\infty$ (provided that we have the scaling
$a^3e^{\lambda_0\vert\phi\vert}\rightarrow 0$, or
$a^3e^{\lambda_0\vert\phi\vert}\rightarrow \infty$ and
$ae^{\lambda_0\vert\phi\vert}\rightarrow 0$, for $a\rightarrow \infty$).
The global wave function for such a case can be compactly written as
\be
\Psi(a, \phi)=K_{\sqrt{-\nu}\over 2}\left({a^2\over 2}\right )
I_{\sqrt{-\nu}\over \vert k\vert}\left(\sqrt{\delta}a^3e^{-{3\sqrt{\g\over
2}\vert\phi\vert}}\right)
\lb{ggg}
\ee

We have therefore found wormholes when $0<\gamma<2/3$. Only when a
cosmological constant is present ($\gamma
=0$) is a quantum wormhole prevented from occurring.

\subsection{Complex scalar field}

The $V(\phi)$ potential when $\gamma>2/3$ suffers
from the presence of a cosmological
constant in its minimum. This prevents a wormhole from occurring since
we require that asymptotically the potential $V(\phi)\rightarrow 0$.

By adding a second scalar field, or equivalently taking a single complex scalar
field, we can ensure that the potential has a zero minimum. This was done
by Ellis, Lyth and Mijic \cite{ell1},
and it will ensure that in the limit $a\rightarrow
\infty$ the potential will also approach zero. One should not think of
this as implying that a complex scalar field is necessary to have wormhole
solutions, but is rather one way of removing  the cosmological constant
caused by a  limitation of the EM procedure in this case.
The detailed discussion of the classical solutions for the case of a
complex scalar field is given in the Appendix.

As we are interested in the case $V\rightarrow 0$, $\g>2/3$,
we will work in the ansatz dominated by the imaginary part of the scalar
field, eq. \rf{f18}, where the potential is given by
\footnote{At this stage we rewrite eq. \rf{f23} explicitly
introducing a modulus in the $\sin$,
to avoid unnecessary imaginary factors in the following
calculations. Obviously, eq. \rf{duno} is the same as eq. \rf{f23}.}
\be
V(\sigma)=\Omega \vert\sin[\lambda\sigma]\vert^{2n}
\label{duno}
\ee
which is plotted for $\g =4/3$ in fig. (6).

The WDW equation is
\be
\left\{a^2{\partial^2\over \partial a^2}+pa{\partial \over \partial a}
+{\partial^2\over \partial \sigma^2}+V(\sigma)a^6-a^4\right\}\Psi
(a, \sigma)= 0
\label{dtre}
\ee
where part of the operator-ordering ambiguities are encoded, as usual, in $p$.

The procedure is now to look for solutions of the WDW equation \rf{dtre},
around the stationary points (minima) of the potential $V(\sigma)$,
which satisfy the HP boundary conditions {\bf (a)}-{\bf (b)} stated in
section (3): these should represent our quantum AE wormholes.

To separate the WDW eq. \rf{dtre}, we follow the same method described in
section (4.1.1).
In particular, we introduce a new variable $\chi$ as
\be
\chi = g(\sigma) a^3
\label{ee1}
\ee
and define the quantity $G$
\be
G^{-1} = \frac{1}{g}\frac{\pr g}{\pr\sigma}
\label{ee2}
\ee
such that the WDW equation \rf{dtre} turns out as
\[
G^2\left[a^2\frac{\pr^2}{\pr a^2} + pa\frac{\pr}{\pr a} -
a^4\right]\Psi(a, \chi) =\]
\be
-\left[ \left(\chi^2\frac{\pr^2}{\pr \chi^2} + (1 - G')\chi \frac{\pr}{\pr
\chi}\right) + a^6VG^2\right]\Psi(a, \chi)
\label{ee5}
\ee
We then follow section (4.1.1) in writing for $g$ the analogous of the
 differential
equation \rf{e26} and compare with solutions of the differential
equation \rf{e27}.

We first note that we have another independent solution of eq. \rf{e27},
which is
\be
y(z) = e^{hz}\sinh^{\frac{1}{1 - \alpha}}{k(\alpha - 1)z}
\label{ee30}
\ee
with the same coefficients $h$ and $k$ as given by eqs.
\rf{e28}-\rf{e29}.
Substitution of formula \rf{ee30} into eq.~\rf{e27} this time gives,
for $\beta =h=0$,
\bea
yy'' & = & y^2[\rho + \alpha k^2\coth^2{k(\alpha - 1)z}]
\non \\
     & = & y^2[k^2 +\frac{\alpha k^2}{\sinh^2{k(\alpha - 1)z}}]
\non \\
     & = & k^2y^{2\al}[\al +y^{2(1-\al)}]\label{e34}
\eea

We consider now the limit in which $y^{2(1-\al)}\ll\al$, or
\be
\sinh^2{k(\alpha - 1)z} \ll \alpha
\label{e35}
\ee
In the limit given by eq. \rf{e35} we can
approximate eq.~\rf{e34} as
\bea
yy'' & \approx & \alpha k^2y^{2\al}
\non \\
    &    =   &  \al k^2\sinh^{2\al\over 1-\al}k(\al-1)z \label{e37}
\eea
We can then identify $y = g(\sigma)$ if $\lambda = k(\alpha - 1)$,
$z=\sigma$ and we impose the conditions $2\al/( 1-\al)=2n$ and
$\al k^2=\Omega/\delta$.
Eq. \rf{ee59} for the parameters $\al, \rho$ and $\delta$ is now replaced by
\bea
\alpha &=& \frac{3\g}{2(3\gamma-1)}
\non \\
\rho &=& \frac{(3\g-2)(3\g-1)}{\gamma}
\non \\
\delta  &=& \frac{2-\g}{6(3\g-1)a^2_0}
\label{e59}
\eea

Finally, we note from eqs. \rf{e39} and \rf{e59} that the condition
\rf{e35} is satisfied if we constrain the scalar field $\sigma$ and the
parameter $\g$ to be in the region
\be
\sigma \simeq 0~~~~~~~,~~~~~~~\g>1/3
\lb{f45}
\ee
which is just the region we are interested in.

{}From the definition of $G$ (eq. \rf{ee2}) and eqs. \rf{ee30}, \rf{ee37},
we can easily see that, in the limit given by eq. \rf{f45},
\be
G^{-2}=k^2\coth^2\lambda\sigma \simeq
k^2\sinh^{-2}\lambda\sigma=k^2\chi^{-2\xi/3}a^{2\xi}\simeq
\frac{1 - G'}{\al G^2}
\label{e44}
\ee
Therefore, for a separation constant $\mu$, we finally obtain
the two equations
\be
\left[a^2\frac{d^2}{d a^2} + pa\frac{d}{d a} -a^4 -\mu
a^{2\xi}\right]\psi(a) = 0
\label{e46}
\ee
and
\be
\left[ \chi^2\frac{\delta^2}{\delta \chi^2}
+ \al\left (\chi{{\delta}\over {\delta\chi}} + \delta\chi^2\right )+{\mu\over
k^2}\chi^{2\xi/3}\right]
\Phi(\chi) = 0
\label{e47}
\ee
We will now consider the solutions of the eqs. \rf{e46}-\rf{e47}
for $\g>2/3$.

\paragraph{~~$\bullet$~~\bold ${\bf 2/3<\g<2}$ }
\indent

\subparagraph{~~$\diamond$~~${a\rightarrow 0}$ }
\indent

We begin by studying the behaviour of the wave function
as $a\rightarrow 0$ and check if it satisfies the boundary condition
{\bf (b)} of HP.

First, as we are working in the range $\g>1/3$ (condition \rf{f45}), we
also have that
\be
4>6(1-\al)
\lb{f50}
\ee
Therefore, as the scale factor goes to zero, we can
neglect the $a^4$ term in eq. \rf{e46}, and the $a$-part of the wave
function can be written in terms of Bessel functions as
\be
\psi(a)\simeq a^{1-p\over 2}\left [c_1K_{1-p\over 2\xi}\left({\sqrt{\mu}
a^{\xi}\over \xi}\right) +c_2I_{1-p\over 2\xi}\left({\sqrt{\mu}
a^{\xi}\over \xi}\right)
\right ],~\mu>0
\lb{f62a}
\ee
or
\be
\psi(a)\simeq a^{1-p\over 2}\left [c_3Y_{1-p\over 2\xi}\left({\sqrt{-\mu}
a^{\xi}\over \xi}\right) +c_4J_{1-p\over
2\xi}\left({\sqrt{-\mu}a^{\xi}\over \xi}\right)
\right ],~\mu<0
\lb{f62b}
\ee

Let us now turn to the $\chi$ part of the solution of the WDW equation.
Remembering that the $\chi$-coordinate in eq.~\rf{ee1} reads
\be
\chi  \simeq  a^3(\lambda\sigma)^{3\over \xi}
\label{e60}
\ee
and noting
that $\xi>0$ for $2/3<\g\leq 2$, this immediately implies that $\chi
\rightarrow 0$, as $a\rightarrow 0$ and $\sigma \simeq 0$.
In the limit $\chi\rightarrow 0$,
since we have that $2>2\xi/3$ for $\g>1/3$ (condition \rf{f45}),
we can neglect the term $\chi^2$ in eq. \rf{e47}, whose approximate solutions
can be again expressed in terms of Bessel functions as
\be
\Phi(\chi)\simeq \chi^{\xi\over 6}\left [d_1Y_{1/2}\left({3\sqrt{\mu}
\over k\xi}\chi^{\xi\over 3}\right) +d_2J_{1/2}\left({3\sqrt{\mu}
\over k\xi}\chi^{\xi\over 3}\right)
\right ]~~~,~~~\mu>0
\lb{f59a}
\ee
or
\be
\Phi(\chi)\simeq
\chi^{\xi\over 6}\left [d_3K_{1/2}\left({3\sqrt{-\mu}
\over k\xi}\chi^{\xi\over 3}\right) +d_4I_{1/2}\left({3\sqrt{-\mu}
\over k\xi}\chi^{\xi\over 3}\right)
\right ]~~~,~~~\mu<0
\lb{f59}
\ee

Finally, combining eqs. \rf{f62a}-\rf{f62b} and \rf{f59a}-\rf{f59} and
using the asymptotic forms of Bessel functions (see, e.g., ref.
\cite{abr}),
we see that the global wave function for $a\rightarrow 0$ is
independent of $\mu$ and behaves as
\be
\Psi(a, \sigma)\simeq
[e_1+e_2a^{1-p}][e_3+e_4(\lambda\sigma) a^{\xi}]
\lb{f66}
\ee
This can be regular and satisfy the HP boundary condition {\bf (b)},
for instance for $e_2=0$.

\subparagraph{~~$\diamond$~~${a\rightarrow \infty}$ }
\indent

We now study the behaviour of the solutions of the WDW equation in the
asymptotic region $a\rightarrow \infty$.
In this case, due to condition \rf{f50}, we can neglect the term $a^{6(1-\al)}$
in the $a$-equation \rf{e46}, and we can write $\psi(a)$ as a combination of
modified Bessel functions $K$ and $I$.
Keeping only the asymptotically bounded solution, we have
\be
\psi(a)\simeq a^{1-p\over 2}K_{1-p\over 4}\left ({a^2\over 2}\right )
\lb{f53}
\ee

For the $\chi$ part of the solution, instead, we have now to distinguish
the cases $\chi \rightarrow 0$ and $\chi \rightarrow \infty$.
In the limit $\chi\rightarrow 0$, $\Phi(\chi)$ is given again
by eqs. \rf{f59a}-\rf{f59}.
Combination of eqs. \rf{f59a}-\rf{f59} and \rf{f53} gives
for the asymptotic global wave function in the region $a\rightarrow
\infty$:
\be
\Psi(a, \sigma)\simeq \sqrt{\pi}a^{-{(1+p)\over 2}}e^{-a^2/2}[e_3+e_4(\lambda
\sigma ) a^{\xi}]
\lb{f61}
\ee
This wave function clearly satisfies the HP boundary condition {\bf (a)}.

Therefore, we conclude that wave functions for the case $2/3<\g\leq 2$
represent quantum AE wormholes in the sense of HP provided that we have
the scaling $a\sigma^{1\over \xi}\rightarrow 0$ as $a\rightarrow \infty$.
\footnote{It can be easily shown that the case $\chi\rightarrow \infty$ is not
AE.}

\section{Discussion and Conclusions}
We have obtained the quantum analogues of the classical CM wormholes. There
are two advantages which make these quantum solutions far
more general than their classical counterparts.
  The requirement that the Ricci tensor has a negative
eigenvalue, in order to produce a wormhole throat,
is not necessary in the quantum wormhole.
In the classical wormhole this requirement was accomplished by
using an asymmetric
analytic continuation
of the matter source. This aspect is included ``automatically'' in the
WDW equation and it results in the avoidance
of singularities as $a\rightarrow 0$.

 Quantum wormholes occur, for example, when ordinary radiation is present.
We saw this using the ``on shell'' potential $V(a)$
which does not give all the possible
 solutions but does enable many of the known quantum wormholes to be
found, e.g. that of a massless real scalar field or a conformally coupled
one. Other quantum wormholes can be found using $V(a)$ for any $\gamma$,
although in general they will not have a simple analytic solution. But,
in analogy to the classical case, there is a CM quantum wormhole for any
value of $\gamma $ in the range $2/3<\gamma\leq2$ .

We next used the potential $V(\phi)$ which allows the full possibilities
of quantum solutions to be found. In this case wormholes were also
obtained when the strong energy condition is violated: something
which is not possible for a classical CM wormhole. Only when a
cosmological constant is present ($\gamma=0$)
we were unable to find a wavefunction that obeys the HP
conditions for a quantum wormhole.  Even then, the second condition {\bf
(b)} regarding
the behaviour as $a\rightarrow 0$ is rather easily satisfied for Planck
sized wormholes. Looking at the WDW equation \rf{dewitt}
one sees that the potential
$V(\phi)$ can be ignored (even if it is constant )
 for $a<1$, and this case behaves like a  massless scalar
field in this limit. One might say that the massless scalar field is
the archetypal wormhole for small scale factors. However, if wormholes
are to play a role in black hole evaporation, it is suspected that
quantum wormholes should be much larger than the Planck size $a\sim 1\sim 10
^{-33}cm$ . It is therefore necessary to try and satisfy condition {\bf
(a)} up to, say, values of $a\sim 10^3$. For a perfect fluid matter
source this was found not to be
possible when the strong energy condition is violated. However, with  a
scalar field, although the matter source was violating the strong
energy condition, wormholes could still be found, except for the case
of a cosmological constant. Such solutions are an entirely quantum
``off shell''
phenomena, since the classical equations allowing us to work with the
classically equivalent potential $V(a)$ can no longer be used.

 It is not necessary to solve the WDW
equation with $V(\phi)$ and $\gamma>2/3$, since  the existence of wormholes
had already been established working with the more
restrictive ``on shell'' potential. However, this could be done provided we
correct for the fact that the potential, which is meant to simulate a
certain value of $\gamma$, fails to do so  at the potential's
minimum.

Because wormholes can be obtained for such a broad range of $\gamma$,
this shows, in agreement with the suggestion of
HP,  that indeed quantum wormholes are a general phenomena
that can  occur for any matter source. Recall that any actual potential
coming from a particle theory, e.g. $V(\phi)=m^2\phi^2$ or
$\lambda\phi^4$,
has an effective $\gamma$ in the range $0<\gamma\simeq2$.
 Only when the potential
behaves like an effective cosmological constant $\gamma\simeq 0$ would
the wormhole be prevented. As the minimum $V(\phi)\rightarrow 0$
is approached, such $\phi^n$ potentials typically have an effective
$\gamma\sim 1$
 and so the wormholes found by HP for a massive scalar field
or $\lambda\phi^4$ can be understood.

We have found that  wormhole states  can be obtained by imposing the
HP boundary condition. In doing so we seem to be losing the possibility
of choosing other wavefunctions which might have other favourable
characteristics. We have in mind  the presence of a Lorentzian regime with
 wavefunctions that represent inflationary behaviour.

 This dilemma can
be understood further by
considering  the problem of caustics in the WDW equation
found by Grishchuk and Rozhansky \cite{caust}, see also
ref. \cite{luk}. In order to have a realistic model with a
Lorentzian regime they required the existence
of a caustic. Below a certain value of the field the caustic does not
develop and the model remains Euclidean. But this aspect (of no caustic)
is exactly what is required to have the possibility of a wormhole. In
other words, the lack of a caustic  was first considered a problem but
is now being used as a  condition for the existence of quantum wormholes.
 This lack of a caustic was considered a potential fault of the
HH boundary condition and indeed there is some relation between this
boundary condition and the wormhole HP one. Typically the HH wavefunction
grows exponentially like $\sim \exp(a^2)$, in contrast to the exponential decay
of the HP one. These two behaviours  were contrasted by Kim \cite{kim}.

 The HP boundary condition also differs in a crucial way
with the Tunneling boundary condition. Whereas the Tunneling boundary
condition is peaked at a large potential $V(\phi)$, which will tend to
produce a Lorentzian regime and inflationary
behaviour for $\gamma<2/3$, the wormhole boundary condition
suggests  a small potential $V(\phi)$. However, the Tunneling boundary
condition
does decay exponentially in an Euclidean region.

      Recently, the wormhole boundary condition has been claimed to be the
more fundamental one \cite{mar},
requiring the dropping or modification of the other
(e.g. HH or Tunneling) boundary conditions.
 This was based on the claim that our universe
is asymptotically flat and this could be a prediction of the wormhole boundary
condition. As our universe is not asymptotically flat
 (it appears FRW) and is also
Lorentzian, other boundary conditions would seem  necessary. Especially if we
required an inflationary phase during the early history of the universe.
\footnote{Some authors do not seem to have a problem with this, saying a
``dynamical'' value for $\Lambda$, unaffected by the wormhole
mechanism, could  occur, cf. ref. \cite{fuk}.}
 We do not therefore think at this stage that the issue of the wormhole
boundary condition can be elevated to a superior status. It is an open
question whether different boundary conditions determined, in some sense,
 by underlying conditions can be
 valid. The so called 3rd quantization \cite{3rdq}
 of the WDW equation might give some
justification for this since more than one type of solution could
be present at once. Otherwise, if there is a
correct universal boundary condition for the WDW equation, it is difficult
to see how the possibility of wormholes is compatible with an earlier
inflationary
regime which presumably caused our universe.

What about the implications for the Coleman
mechanism for the setting to zero of the cosmological constant? In an
earlier version of this paper \cite{cou2}
we had only considered the $V(a)$ potential
and had missed seeing the possibility of having wormholes when the strong
energy condition is violated. We had argued that there was a contradiction
in that wormholes are incompatible with a $\Lambda$ term. We then notice
that, as mentioned in ref. \cite{3rdq},  the wormhole mechanism does first
require $\Lambda<<1$, this requirement presumably coming
from some other mechanism,
e.g. supersymmetry. Once $\Lambda$ attains this ``small'' value, wormholes
would then be possible and could proceed in setting $\Lambda$
infinitesimally close to zero \cite{col}.
Recall that we are trying to explain the current value
of $\Lambda\sim 10^{-120}$.

However, we are still left with the problem of how the boundary
conditions responsible for wormholes are compatible with Lorentzian
or inflationary behaviour. A related problem is that: if the Tunneling
boundary condition and wormholes could coexist, the Coleman mechanism would
work in the opposite direction, setting $\Lambda\sim 1$, i.e. large
\cite{cou3}.

Note that while the boundary conditions that enable the Coleman mechanism
to proceed, given wormholes, are fairly general (the Tunneling
boundary condition
is an exception) we have found that obtaining  wormholes themselves is
a more restrictive requirement. Restrictive, in the
sense that they might preclude other things from occurring,
e.g. inflation. Within the Coleman mechanism,
it might be preferable to try and do without wormholes {\em per se},
and rather other things, e.g. axions, torsion \cite{sabb},
could simulate their effects. We would then not have to impose the
restrictive wormhole boundary condition, but rather concentrate on the
boundary conditions
explaining our large Lorentzian universe with its possible past
inflationary epoch.

\vspace{33pt}
\noindent {\Large \bf Acknowledgements}{\vspace{11pt}}

We would like to thank Prof. G.F.R. Ellis for very helpful comments and
suggestions during the course of this work.
We should also thank Dr. L.P. Garay for an interesting e-mail exchange.
A.C. is supported by an individual EEC fellowship in the `Human Capital
and Mobility' program, under contract No. ERBCHBICT930313.
\vspace{33pt}

\appendix
\section{Appendix}
\subsection{Classical model for a complex scalar field}

Consider the model with a single complex scalar field $\varphi$ and
Lorentzian action
\footnote{Note here the `unusual' definition of the kinetic term
for the scalar, instead of the standard $g^{\mu\nu}\pr_{\mu}\varphi\pr_{\nu}
\varphi^{\ast}$.}
\be
S_M=-{1\over 2}\int d^4x\sqrt{-g}[g^{\mu\nu}\pr_{\mu}\varphi\pr_{\nu}
\varphi+V(\varphi)]
\lb{f1}
\ee
The complex scalar can be defined as
\be
\lambda\varphi = \lambda\phi + i[\lambda\sigma - (s +
\F{1}{2})\pi]\dot=\lambda\phi + i\lambda\tilde\sigma
\lb{I70}
\ee
where $\lambda$ is defined in eq. \rf{param}
and both $\phi$ and $\sigma$ are real (and $s$ is an integer).
Using eq. \rf{I70} and the homogeneous and isotropic
FRW metric ansatz \rf{x},
it is then immediate to explicitly write down the Lorentzian action (for a
homogeneous scalar $\varphi$) as
\be
S=-{1\over 2}\int d\tau a^{(3\g-2)/2}[\dot a^2-(\dot
\phi^2+2i\dot\phi\dot\sigma-\dot\sigma^2)a^2+a^{3(2-\g)}V(\varphi)
-a^{4-3\g}]
\lb{f3}
\ee
Following the standard procedure we
can then define the energy-momentum tensor of the scalar field as
\bea
T_{\mu\nu}&=&\pr_{\mu}\varphi\pr_{\nu}\varphi-{g_{\mu\nu}\over 2}
[\pr_{\rho}\varphi\pr^{\rho}\varphi+V(\varphi)]
\non \\
&\dot =&pg_{\mu\nu}+(p+\rho)U_{\mu}U_{\nu}
\lb{f4}
\eea
Here we have imposed that, in some appropriate limit, the scalar field
can be approximated by a perfect fluid with pressure $p$, energy
density $\rho$, 4-velocity vector $U^{\mu}$ and
where (see, for instance, ref. \cite{ell2})
\bea
\pr_{\mu}\varphi &=&-{\dot\varphi\over a^{(4-3\g)/2}}U_{\mu}
\non \\
U_{\mu}U^{\mu}&=&-1
\lb{f5}
\eea
Substituting eq. \rf{f5} in eq. \rf{f4} we can easily express the
energy density and pressure of the `fluid' as a function of the
scalar field content of the theory, i.e.
\bea
\rho&=& {1\over 2}\left [{(\dot \phi^2+2i\dot\phi\dot\sigma-\dot\sigma^2)
\over a^{4-3\g}}+V(\varphi)\right]
\non \\
p&=& {1\over 2}\left [{(\dot \phi^2+2i\dot\phi\dot\sigma-\dot\sigma^2)
\over a^{4-3\g}}-V(\varphi)\right]
\lb{f6}
\eea
Then, as done in ref. \cite{cm}, we can work in the restricted ansatz
described by the equation of state
\be
p=(\g-1)\rho
\lb{f7}
\ee
The idea is now to consider the classical
evolution of the scalar field model in the limits in which the real
component ($\phi$) or the imaginary component ($\sigma$) respectively
dominate the equations of motion.

Let us first consider the ansatz
\be
\phi\gg\tilde \sigma
\lb{f9}
\ee
The Friedmann equation for which the action \rf{f3} is stationary is,
in this limit,
\be
-\dot a^2+\dot \phi^2a^2+a^{3(2-\g)}V(\varphi)-a^{4-3\g}\simeq 0
\lb{f10}
\ee
Using the method of ref. \cite{em}, we assume that the Lorentzian geometry
\rf{x} represents an expanding universe with scale factor
\be
a=a_0\left [1-\left({3\g-2\over 2}\right )^2{\tau^2\over a_0^{3\g-2}}
\right ]^{1/(3\g-2)}
\lb{f11}
\ee
and find the matter content which is necessary to drive such an
instanton.
Using eqs. \rf{f6} and \rf{f7}, we find that
\be
V(\varphi)\simeq {(2-\g)\over \g}{\dot \phi^2\over a^{4-3\g}}
\lb{f13}
\ee
Eliminating $V$ by means of eq. \rf{f13} and solving for $\phi$
in eq. \rf{f10}, we obtain
\be
\lambda(\phi-\phi_0)= ~\mbox{arctanh}
\left [{(3\g-2)\tau\over 2a_0^{(3\g-2)\over 2}}\right]
\lb{fg}
\ee
Setting $\phi_0=0$, the form of
the classical potential for the scalar field is
\be
V(\varphi)\simeq{(2-\g)\over 2a_0^2}\cosh^{6\g\over 3\g-2}\lambda\phi
\lb{f15}
\ee
(the special case $\g=0$ corresponds to the constant potential
$V(\varphi)=1/ a_0^2$).
This is the CM solution.

Let us now consider the ansatz in which the scalar field $\varphi$
is dominated by the imaginary part $\sigma$, i.e.,
\be
\tilde \sigma \gg \phi
\lb{f18}
\ee
and analytically continue to the Euclidean region by Wick
rotating the time according to $\tau\rightarrow i\tau$.
The action in the Euclidean region becomes
\be
S_E={1\over 2}\int d\tau a^{(3\g-2)/2}[\dot a^2+
\dot\sigma^2a^2-a^{3(2-\g)}V(\varphi)
+a^{4-3\g}]
\lb{f17}
\ee
The equations of motion which one can derive from the action \rf{f17}
from variations of the lapse and of the scale factor are, respectively,
\bea
0&\simeq &\dot a^2+\dot \sigma^2a^2+a^{3(2-\g)}V(\varphi)-a^{4-3\g}
\non \\
0&\simeq &(3\g-2)\dot a^2+4a\ddot a+(3\g+2)a^2\dot \sigma^2
\non \\
&+&(10-3\g)V(\varphi)a^{3(2-\g)}-3(2-\g)a^{4-3\g}
\lb{f19}
\eea
We then assume that the Euclidean geometry (at least asymptotically)
is that of an AE wormhole,
\be
a=a_0\left [1+\left({3\g-2\over 2}\right )^2{\tau^2\over a_0^{3\g-2}}
\right ]^{1/(3\g-2)}
\lb{f20}
\ee
Eliminating $\dot\sigma$ from eqs. \rf{f19} then gives, as in the
previous case,
\be
V(\varphi)\simeq {(2-\g)\over \g}{\dot \sigma^2\over a^{4-3\g}}
\lb{f21}
\ee
and using again this formula in the first of eqs. \rf{f19} allows
to solve for the field $\sigma$ as
\be
\lambda(\sigma-\sigma_0)=
\mbox{arccotan}\left[{(3\g-2)\tau\over 2a_0^{(3\g-2)\over 2}}\right]
\lb{gf}
\ee
Choosing $\sigma_0=0$,
finally gives for the classical potential of the scalar field
\be
V(\varphi)\simeq{(2-\g)\over 2a_0^2}\sin^{6\g\over 3\g-2}\lambda\sigma
\lb{f23}
\ee
(as before, the special case $\g=0$ corresponds to the constant potential
$V(\varphi)=1/ a_0^2$).
It is now easy to show that the
two `asymptotic' forms of the scalar field potential given by eqs.
\rf{f15} and \rf{f23} can be both derived from a single, semipositive
definite potential of the form
\be
V(\varphi) = \Omega \left[ \X{[\lambda\varphi]}\X{[\lambda\varphi^*]}\right]^n
\lb{I73}
\ee
where $n$ and $\Omega$ are defined by eq. \rf{param}.
Using eq. \rf{I70}, this potential can be also written as an explicit function
of $\phi$ and  $\sigma$ as
\be
V(\varphi) = \Omega\5[\cosh^2{[\lambda\phi]} - \cos^2{[\lambda\sigma]}\6]^{n}
\lb{II73}
\ee
It is also clear that the form \rf{f15} of the potential
is obtained in the limit of large $\phi$, while the
form \rf{f23} corresponds to the limit of small $\phi$,
in agreement with the previous hypothesis.
Therefore, the Friedmann equations in the two asymptotic regions will
read (eliminating $\dot \phi^2$ and $\dot \sigma^2$ by means of eqs.
\rf{f13} and \rf{f21} from, respectively, eqs. \rf{f10} and
\rf{f19})
\be
H^2 + a^{-2} \simeq a^{-2}_0\cosh^{2n}
{[\lambda\phi]}
\lb{f24a}
\ee
\be
H^2+a^{-2}\simeq a^{-2}_0\sin^{2n}{[\lambda\sigma]}
\lb{f24b}
\ee

At this point it should be mentioned that a similar behaviour of
the potential could have been obtained also by using one single
real scalar field according to the lines of ref. \cite{c1}.
In this case, however, if one also wants to impose smooth matching
conditions for the potential and kinetic energy of the scalar field
at the throat of the wormhole, an extra Wick rotation of the scalar
field itself when passing from the Lorentzian to the Euclidean
region is required.
Eqs. \rf{f15} and \rf{f23} now represent the Lorentzian and Euclidean
form of the same potential.

\newpage
\noindent {\Large \bf Figure Captions}
\bigskip
\bigskip
\begin{description}
\item[Fig. 1] The wave function for a perfect fluid model with $\g=4/3$
(we have plotted the sum in eq. \rf{f78a} for $n\in [0, 10]$ and with $a_0=1$).
\item[Fig. 2] The wave function for a perfect fluid model with $\g=0$
(we have plotted eq. \rf{f77a} for $b_2=0$ and $a_0=1$).
\item[Fig. 3] The potential $V(\phi)=\Omega\cosh^{2n}\lambda\phi$ for
$\g=1/2$.
\item[Fig. 4] The potential $V(\phi)=\Omega\cosh^{2n}\lambda\phi$ for
$\g=4/3$.
\item[Fig. 5] The potential $V(\phi)=\Omega\cosh^{2n}\lambda\phi$ for
$\g=0$.
\item[Fig. 6] The potential $V(\phi)=\Omega\sin^{2n}\lambda\sigma$ for
$\g=4/3$.
\end{description}

\end{document}